\documentclass[12pt,a4paper]{article}
\usepackage{geometry}
\usepackage{graphicx}
\usepackage{amsmath}
\usepackage{amssymb}
\usepackage{epstopdf}
\usepackage{hyperref}
\usepackage{lipsum}
\usepackage{chngpage}
\usepackage{float}
\usepackage{subfigure}
\usepackage{color,pictexwd,psfrag,pstricks}
\restylefloat{table}
\usepackage{dcolumn}
\usepackage{bm}
\usepackage{tikz}
\newcommand{\figs}{draft_fig/}
\newcommand{\ddxi}[2]{\frac{\partial{#1}}{\partial{#2}}}

\definecolor{red}{rgb}{1.0,0.0,0.0}
\definecolor{green}{rgb}{0.0,1.0,0.0}
\definecolor{blue}{rgb}{0.0,0.0,1.0} 

\newcommand{\redline}{\raisebox{2pt}{\tikz{\draw[-,red,solid,line width = 0.9pt](0,0) -- (5mm,0);}}}
\newcommand{\redlinedash}{\raisebox{2pt}{\tikz{\draw[-,red,dashed,line width = 0.9pt](0,0) -- (5mm,0);}}}
\newcommand{\greenline}{\raisebox{2pt}{\tikz{\draw[-,black!40!green,solid,line width = 0.9pt](0,0) -- (5mm,0);}}}
\newcommand{\blueline}{\raisebox{2pt}{\tikz{\draw[-,blue,solid,line width = 0.9pt](0,0) -- (5mm,0);}}}
\newcommand{\greyline}{\raisebox{2pt}{\tikz{\draw[-,gray,solid,line width = 0.9pt](0,0) -- (5mm,0);}}}
\newcommand{\cyanline}{\raisebox{2pt}{\tikz{\draw[-,cyan,solid,line width = 0.9pt](0,0) -- (5mm,0);}}}
\newcommand{\cyanlinedash}{\raisebox{2pt}{\tikz{\draw[-,cyan,dashed,line width = 0.9pt](0,0) -- (5mm,0);}}}
\newcommand{\purpleline}{\raisebox{2pt}{\tikz{\draw[-,red!40!blue,solid,line width = 0.9pt](0,0) -- (5mm,0);}}}

\newcommand{\yellowline}{\raisebox{2pt}{\tikz{\draw[-,yellow!80!red,solid,line width = 0.9pt](0,0) -- (5mm,0);}}}

\begin{document}

\begin{center}

\textbf{\LARGE{Neural network-based modelling of unresolved stresses in a turbulent reacting flow with mean shear.}}

\vspace{0.25cm}

Z.M. Nikolaou \footnote[1]{Computation-based Science and Technology Research Centre (CaSToRC), The Cyprus Institute, Nicosia, 2121, Cyprus. z.nicolaou@cyi.ac.cy}, C. Chrysostomou \footnotemark, Y. Minamoto \footnote[2]{Department of Mechanical Engineering, Tokyo Institute of Technology, 2-12-1 Ookayama, Meguro-ku, Tokyo, 152-8550, Japan.}, L. Vervisch \footnote[3]{CORIA - CNRS, Normandie Universit\'e, INSA de Rouen Normandie, France.} 

\end{center}

\begin{center}

\textbf{Abstract}

\end{center}

\noindent Data-driven methods for modelling purposes in fluid mechanics are a promising alternative given the continuous increase of both computational power and data-storage capabilities. Highly non-linear flows including turbulence and reaction are challenging to model, and accurate closures for the unresolved terms in large eddy simulations of such flows are difficult to obtain. In this study, we investigate the use of artificial neural networks for modelling an important unclosed term namely the unresolved stress tensor, in a highly demanding turbulent and reacting flow, which additionally includes mean shear. The performance of the neural network-based modelling approach is conducted a priori following a coarsened mesh approach, and compared against the predictions of eight other classic models in the literature, which include both static and dynamic formulations. 

\newpage

\section{Introduction}

Large Eddy Simulation (LES) is a powerful modelling tool for the simulation of turbulent and reacting flows. LES reduces the computational load substantially in comparison to Direct Numerical Simulation (DNS), by resolving only the largest, energy-containing motions of the flow. Although a typical LES is computationally more expensive than a Reynolds Averaged Navier Stokes (RANS) simulation, the increase in computational power in recent years has established LES as the de facto modelling approach in many industries for simulating fluid flows in devices with complex geometries and of a realistic size \cite{pitsch_annrev_2006,gicquel_pec_2012}. An important ingredient in LES, is the set of underlying closures used to model the unresolved terms which appear in the spatially-filtered transport equations. 

The unresolved stress tensor in particular in the filtered momentum equation, $\tau_{ij}$, determines the dissipation/backscatter of kinetic energy as a result of unresolved motions \cite{sagaut_book_2001}. This term affects the evolution of important bulk flow quantities, and in turn the evolution of the flow field. Numerous models have been developed in the literature throughout the years for this important term, mainly aimed at incompressible and non-reacting flows-a detailed review is given in \cite{meneveau_annrev_2000}. In the standard modelling approach, the unresolved stress tensor is modelled using properly ``tuned" algebraic functions of the resolved quantities on the mesh. These, typically include the filtered velocities $\bar{u}_i$ (as well as higher-order filtered values) and their spatial gradients $\partial{\bar{u}_i}/\partial{x_j}$. Some of the most popular models include eddy-diffusivity models such as the static/dynamic Smagorinsky \cite{smagorinsky_mwr_1963,germano_pof_1991,moin_pof_1991}, scale-similarity models \cite{bardina_tf19_1983,vreman_tcfd_1996}, models of the gradient type \cite{clark_jfm_1979}, and mixed models \cite{zhang_pof_1993,salvetti_pof_1994,nicoud_ftc_1999,lodato_pof_2009,wincelmans_pof_2001}. Such models have been widely used in LES of both non-reacting and reacting flows, with varying levels of success. The majority of these models are relatively straightforward to implement, while the computational cost can vary significantly especially if a model involves the evaluation of dynamic parameters. A common characteristic for the majority of such models is that they usually involve some simplifying assumption in their development, which may not be valid for conditions other than those originally developed for. For example, the assumption of the unresolved stresses being aligned with the rate of strain tensor in the Smagorinsky model, is a rather strong one. Previous theoretical as well as experimental work showed this assumption to be invalid both for non-reacting \cite{tao_pof_2000,tao_jfm_2002} and for reacting flows \cite{pfandler_expfluids_2010,klein_caf_2015}. Furthermore, the majority of classic models involve tunable parameters whose spatio-temporal value depends on the flow regime and on the reaction mode. As a result, a single universal method for accurate parameterision/regularisation of the models' costants is difficult to obtain. All of the above, limit the predictive ability of LES only to conditions where the models for the unresolved terms are known to perform well.   

An alternative promising modelling approach is based on machine-learning techniques. Machine-learning methods such as Artificial Neural Networks (ANNs) and Convolutional Neural Networks (CNNs) have been widely used to solve classification and regression problems (amongst other) in image recognition \cite{krizhevsky_procneural_2012}, text translation \cite{sutskever_procadvneur_2014}, decision making \cite{Mnih_nat_2015,silver_nat_2016}, gene-profiling \cite{Khan_nat_2001} etc. by directly exploiting the abundance of information contained within very large data sets. In the fluid mechanics community, DNS databases of a range of non-reacting flows are of the order of petabytes \cite{kanov_compsceng_2015}. In reacting flows, simulations using DNS with detailed chemistry and multi-step reduced chemistry are becoming more common \cite{minamoto_pof_2011,nikolaou_cnf_2014,nikolaou_cst_2015,aspden_cnf_2016,wang_jfm_2017}, while numerical solvers are being developed for DNS aimed at the exascale \cite{jchen_exapp_2017} and exploiting hybrid architectures \cite{hernandez_caf_2018}. As a result, the application of machine-learning techniques using data from high-fidelity simulations for modelling purposes in LES is a timely one \cite{kutz_jfm_2017}. In the seminal work of Hornik \cite{hornik_nn_1991} it was shown that a feedforward neural network, even with a single hidden layer, is a universal function approximator in the limit of a sufficiently large number of nodes in the hidden layer. As result, algebraic closures of increased orders of complexity can in principle be developed for the unresolved stress tensor by adjusting the number of layers and/or nodes of a neural network. Some common techniques which are widely used for the analysis/modelling of fluid flows can thus be casted in a neural-network framework. For example, it was shown in \cite{milano_jcp_2002} that a Proper Orthogonal Decomposition (POD) is essentially a subset of a more general non-linear ANN. In the same study \cite{milano_jcp_2002}, a neural network was developed so as to reconstruct the near-wall flow, in the context of RANS, which outperformed the results obtained using a POD-based model. 

Despite the abundance of data in the fluid mechanics community, neural networks have been primarily employed for control purposes, while the literature for modelling purposes is much more limited. In the context of RANS, a neural network was trained in \cite{ling_jfm_2016} to predict the anisotropy stress tensor in shear flows, from knowledge of the mean strain rate tensor, and the mean rotation rate tensor. In \cite{tracy_aiaa_2015} a neural network was trained, using results from RANS simulations, to reproduce the results obtained using the Spallart-Almaras model. A closely related method, using random forests, was employed in \cite{wang_prf_2017} in order to essentially calibrate regression functions for the discrepancy between RANS-modelled and DNS-modelled flow variables, which was used to improve the predictive ability of the RANS simulations. In \cite{singh_AIAA_2017} experimental data instead of simulation data were used, in order to train a neural network to augment the predictive capability of the Spallart-Almaras model in RANS simulations of airfoil flows. In \cite{ma_pof_2015} a neural network was used to directly model the ``streaming" stresses in vertical two-phase flows. In the context of LES, a neural network was used in \cite{sarghini_caf_2003} to calculate the dynamic parameter of the eddy-diffusivity component in a mixed stress tensor model, in turbulent channel flow. In \cite{moreau_pof_2006} a neural network was trained in order to develop optimal estimators for the unresolved scalar variance, by examining in a structured manner different sets of inputs, a process which would otherwise be difficult to replicate using classic algebraic models. In \cite{gamahara_prf_2017} a neural network was trained, using data from direct simulations, to directly predict the unresolved stresses from the resolved flow quantities in LES of turbulent channel flow.  An a posteriori LES study was also conducted with overall good results \cite{gamahara_prf_2017}. A slightly different modelling approach based on deconvolution was examined in \cite{maulik_jfm_2017} for incompressible turbulence: a network was trained to deconvolute the filtered velocity components, and the unresolved stress tensor was modelled by explicitly filtering the deconvoluted velocity fields. In a more recent study \cite{wang_pof_2018}, an ANN was trained to directly predict $\tau_{ij}$ from the resolved flow variables, for incompressible homogeneous decaying turbulence, and an posteriori LES study was also conducted with the ANN-based modelling framework outperforming the results obtained using a dynamic Smagorinsky model.

In the context of reacting flows, machine-learning techniques have been primarily used for modelling the chemical kinetics \cite{ihme_proccomb_2009,sen_proccomb_2009,sen_cnf_2010}. 
For modelling unresolved terms, convolutional networks were successfully employed in \cite{lapeyre_arxive_2018} to model the flame surface density, and in \cite{nikolaou_arxive_2018,nikolaou_ftc_2019} for modelling the progress variable variance in a deconvolution-based context. As for the stress tensor, the standard approach is to employ models originally developed and validated for incompressible and non-reacting flows. As indicated however in many previous theoretical and experimental studies, the effects of heat release are not properly accounted for in such models \cite{klein_caf_2015,bray_cst_1981,chomiak_cnf_1995}. This important point was also noted in \cite{klein_caf_2015} where a number of different stress-tensor models were evaluated for turbulent premixed freely-propagating flames. It was additionally shown in \cite{klein_caf_2015} that the standard averaging procedure for regularising the dynamic parameters e.g. $C_D$ in the Smagorinsky model, was inadequate in the case of reacting flows. In light of these issues, Schoepplein et al. \cite{schoepplein_jcp_2018} has recently employed a data-driven method namely Gene-Expression Programming (GEP), in order to derive best-fit functional relationships for the unresolved stress-tensor in LES of turbulent premixed flames. 

In this study, a high-fidelity DNS database of a demanding multi-physics flow which includes turbulence, reaction, and mean shear, is used in order to train an artificial neural network to predict all six individual components of the unresolved stress tensor $\tau_{ij}$. The nature of the flow presents a challenging testing case for any stress tensor model. Furtheremore, in contrast to most a priori studies in the literature, the evaluation of the method is conducted following a simulated LES approach where the effects of mesh coarsening are taken into account \cite{nikolaou_ftc_2018,nikolaou_prf_2018}. It is important to note at this point that even though this constitutes an a priori study, it forms an important first step for developing models for unresolved terms in LES. Furthermore, previous successful a posteriori testing using models developed from a priori studies indicates the merit of this approach \cite{wang_pof_2018}.    

\section{Direct numerical simulation database}\label{sec:dns}

The fully compressible governing equations for the conservation of mass, momentum, temperature, and species mass fractions are solved for, using an in-house DNS code (TTX), 

\begin{equation}
\frac{\partial \rho}{\partial t}+\nabla\cdot\left(\rho\boldsymbol{u}\right)=0,
\end{equation}

\begin{equation}
\frac{\partial\left(\rho\boldsymbol{u}\right)}{\partial t}+\nabla\cdot\left(\rho \boldsymbol{uu}\right)=-\nabla\cdot \boldsymbol{P},
\end{equation}

\begin{eqnarray}
\frac{\partial\left(\rho T\right)}{\partial t}
+\nabla\cdot\left(\rho\boldsymbol{u}T\right)
=\frac{1}{\bar{c}_v}\nabla\cdot\left(\lambda_{th}\nabla T\right)
-\frac{1}{\bar{c}_v}\sum_{i=1}^{N}\left(\rho Y_i\boldsymbol{V}_i c_{p,i}\cdot\nabla T\right)\nonumber\\
-\frac{T}{\bar{c}_v}\sum_{i=1}^{N}\left[R_i\nabla\cdot\left(\rho Y_i\boldsymbol{V}_i\right)\right]
-\frac{1}{\bar{c}_v}\boldsymbol{P}:\left(\nabla\boldsymbol{u}\right)\nonumber\\
-\frac{1}{\bar{c}_v}\sum_{i=1}^{N}\left(h_i \omega_i\right)+\frac{T}{\bar{c}_v}\sum_{i=1}^{N}\left(R_i \omega_i\right),
\end{eqnarray}

\begin{equation}
\frac{\partial\left(\rho Y_i\right)}{\partial t}+\nabla\cdot\left(\rho\boldsymbol{u}Y_i\right)=-\nabla\cdot\left(\rho Y_i \boldsymbol{V_i}\right)+\omega_i,
\end{equation}

\noindent where $\lambda_{th}$, $h_i$, $R_i$ and $\omega_i$ denote the mixture's thermal conductivity, specific enthalpy, gas constant, and reaction rate of species $i$, respectively. The stress tensor $\boldsymbol{P}$ is given by,
\begin{equation}
\boldsymbol{P}=\left[p+\frac{2}{3}\mu\left(\nabla\cdot\boldsymbol{u}\right)\right]\boldsymbol{I}-\mu\left[\left(\nabla \boldsymbol{u}\right)+\left(\nabla \boldsymbol{u}\right)^T\right],
\end{equation}
\noindent where $\mu$ is the dynamic viscosity of the mixture. Fickian-type diffusion is used for modelling the species' diffusion velocities $\boldsymbol{V_i}$. The governing equations are discretised on a uniform mesh, using a fourth-order central finite-difference scheme for the spatial derivatives, and a third order Runge-Kutta scheme for the time derivatives. The chemical source terms are integrated using a point-implicit method to reduce stiffness. The adiabatic combustion of a stoichiometric hydrogen-air mixture at 0.1 MPa is simulated using a detailed chemical kinetic mechanism with 27 reactions and 12 species namely H$_2$, O$_2$, H$_2$O, O, H, OH, HO$_2$, H$_2$O$_2$, N$_2$, N, NO$_2$ and NO \cite{Gutheil1993}. The temperature dependence of viscosity, thermal conductivity and diffusion coefficients are calculated using the CHEMKIN-II packages \cite{kee_sandia_86, kee_sandia_89}. Soret, Dufour, pressure gradient and radiative heat transfer effects are neglected. 

\begin{figure}[h!]
\subfigure[]{
\includegraphics[width=0.40\textwidth, trim=2.0cm 0.0cm 0.0cm 0.0cm]{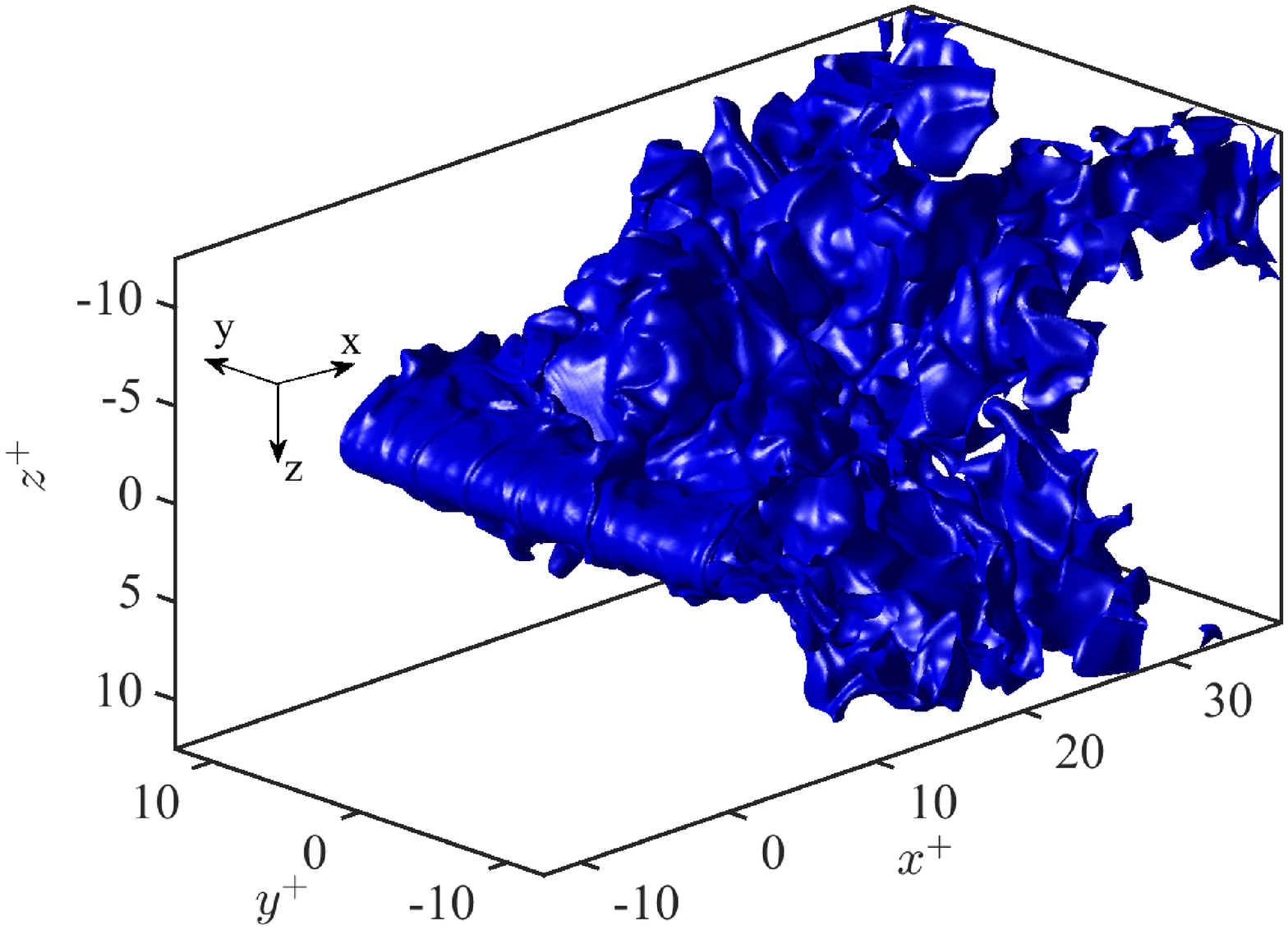}
}
\subfigure[]{
\includegraphics[width=0.40\textwidth, trim=2.0cm 0.0cm 0.0cm 0.0cm]{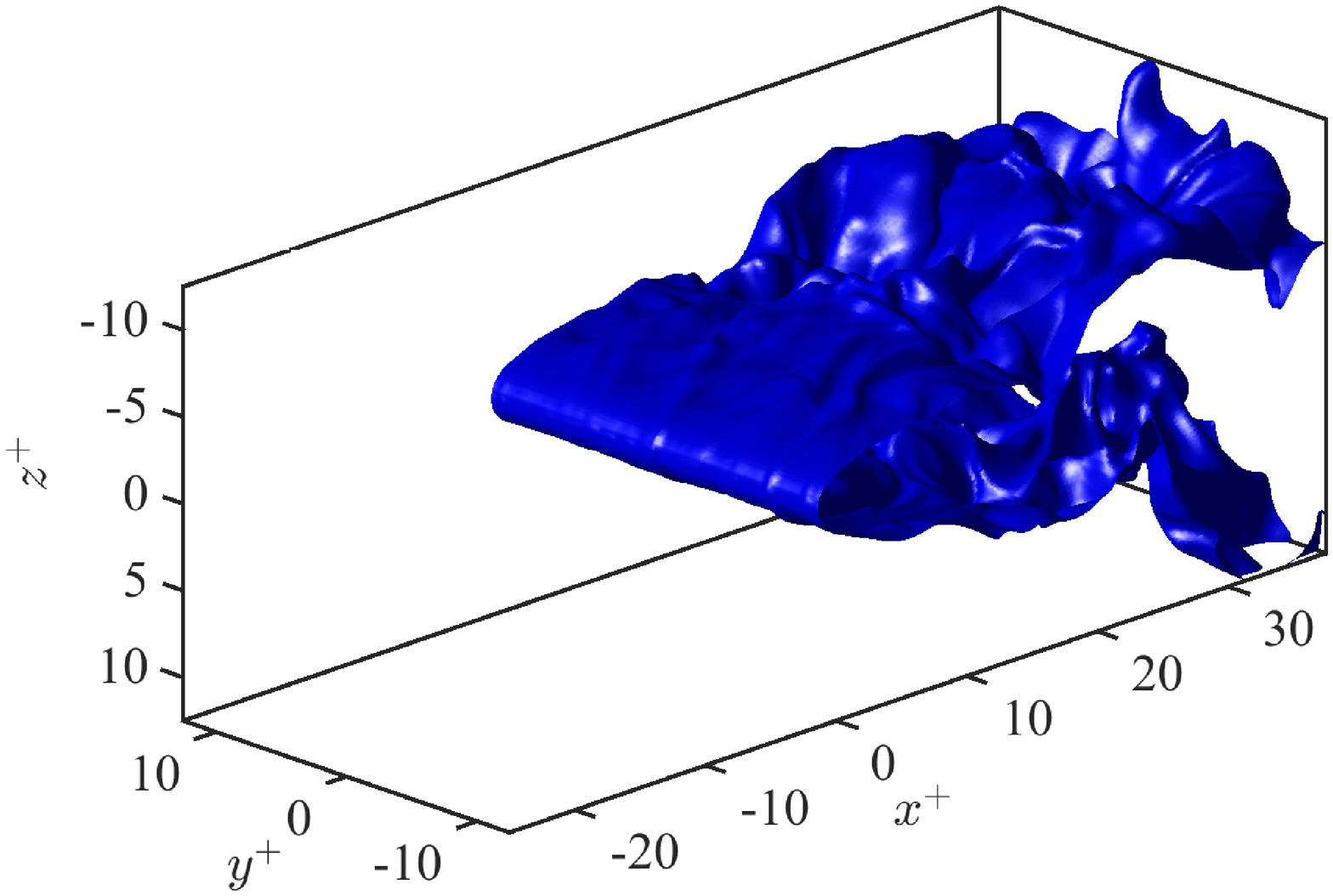}
}
\subfigure[]{
\includegraphics[width=0.40\textwidth, trim=2.0cm 0.0cm 0.0cm 0.0cm]{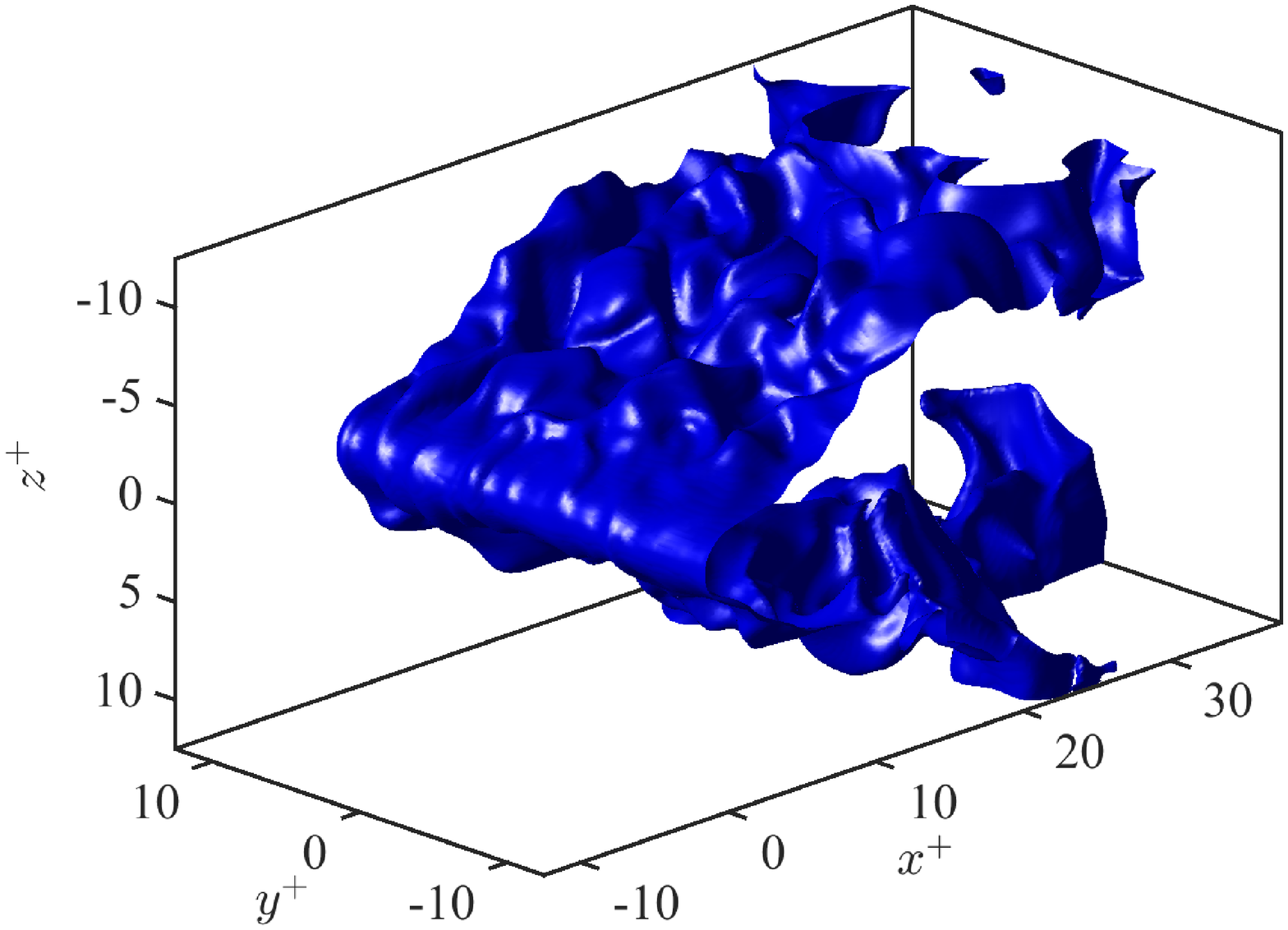}
}
\caption{Progress variable iso-surface $c=0.1$, for (a) case V97, (b) case V60H, and (c) case V60.}
\label{fig:c_01_iso}
\end{figure}

The computational domain and the coordinate system are shown in Fig. \ref{fig:c_01_iso}. Inflow-outflow boundaries are used in the $x$ direction, outflow boundaries are applied in the $z$ direction, and a periodic boundary is specified in the $y$ (homogeneous) direction. From the inflow $x$-boundary, the premixed unburnt mixture and turbulence are fed at a mean inlet velocity $\bar{u}_{in}$. The inflowing turbulence is obtained by first performing DNS of incompressible homogeneous isotropic turbulence using a spectral method. In the present DNS, a no-slip flame anchor (rod) consisting of a high temperature ($\sim2000$~K) region is placed at a distance $x_{r}$ from the inflow boundary. The temperature inside the flame anchor is based on a Gaussian distribution and its diameter $d\simeq\delta_{th}$, where $\delta_{th}$ is the flame thermal thickness (0.49~mm) \cite{minamoto_pof_2011}. Also, in the initial field, the temperature is gradually decreased to match the unburnt gas temperature. Once the simulation has started, the mixture naturally starts reacting and a turbulent V-flame is established-Figure \ref{fig:c_01_iso} shows an instantaneous progress variable iso-surface based on the hydrogen mass fraction, $c=(Y_{H_2}-Y^r_{H_2})/(Y^p_{H_2}-Y^r_{H_2})$. Note that the axes are normalised ($^+$) using the laminar flame thickness $\delta_L$ (0.2~mm) and are relative to the rod location. After the initialization, the DNS is run for three flow-through times, $L_x/\bar{u}_{in}$, to ensure that any initial transients have been evacuated. Three V-flames are considered in the present study, and their conditions are summarised in Table~\ref{tbl:turb_param}. $u_{rms}$ is the root-mean-square value of the fluctuating component of the incoming turbulence field, with an integral length scale $l_T$. The turbulence Reynolds number is $Re_T=u_{rms}l_{T}/ \nu_r$, the Damkohler number is $Da=(l_T /u_{rms})/(\delta / s_{L})$ and the Karlovitz number is $Ka=(\delta/ \eta _k)^2$, where $s_{L}$ is the laminar flame speed, and the diffusive thickness $\delta=\nu_r/s_{L}$. The laminar flame thickness is defined as $\delta _L$=$(Y^r_{H_2}-Y^p_{H_2})/\max(dY_{H_2}/dx)$ where $Y_{H_2}$ is the hydrogen mass fraction profile obtained from a laminar flame calculation.

\begin{figure}[h!]
\centering
\includegraphics[scale=0.45, trim=5.0cm 0.0cm 5.0cm 0.0cm]{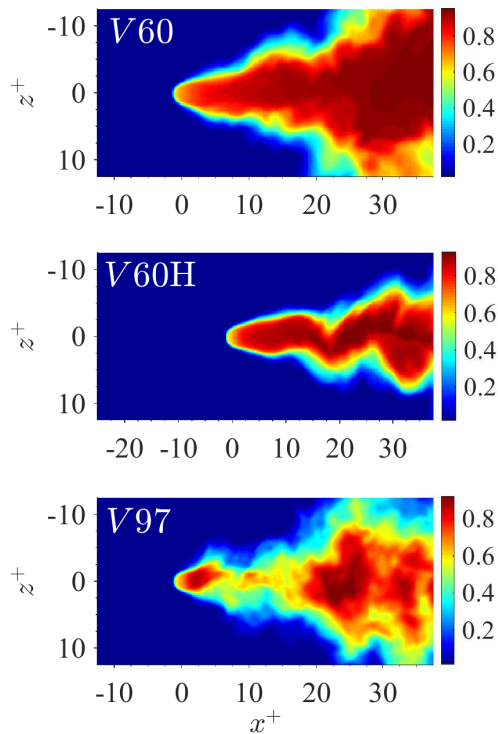}
\caption{Averaged (in homogeneous $y$ direction) instantaneous progress variable, $<c>$, for all three cases.}
\label{fig:c_samp_avr_allcases}
\end{figure}

The computational domain size, $L_x\times L_y\times L_z$, is $10\times 5\times 5$~mm for V60 and V97 and $12.5\times 5\times 5$ mm for V60H. The rod is placed at $x_r$=2.5 mm for cases V60 and V97, and at $x_r$=5.0 mm for case V60H. These domains are discretised on a uniform mesh $N_x\times N_y\times N_z$ of $513\times 257\times 257$ points for V60, $641\times 257\times 257$ points for V60H, and $769\times 385\times 385$ points for V97. These resolutions ensure that there are at least 20 mesh points inside $\delta_{th}$ so that it is well resolved. Also, the resolution for the DNS was found to be more than sufficient to resolve turbulence and the boundary layers near the flame anchor. Figure \ref{fig:c_samp_avr_allcases} shows instantaneous $y$-averaged distributions of the progress variable $c$ for the three different cases. Note the extinction events taking place for case V97 immediately downstream of the rod due to the excessive shear and relatively higher turbulence level. This is the most demanding case of the three in terms of modelling the unresolved stress tensor and has been used for testing the neural network. 

\begin{table}
\centering
\begin{tabular}{l c c c c c c c c c c}
\hline
\hline
 Case &$\bar{u}_{in}/s_L$ & $u_{rms}/s_L$ & $l_{T}/{\delta}$ & $Ret$ & $Da$ & $Ka$ & $L_x$ (mm) & $L_y$ (mm) & $L_z$ (mm) & $x_r$ (mm)\\ [0.75ex]
 \hline
 V60  &10.0  &2.2  &100.0 &220.0  &45.5 &0.33  & 10.0 & 5.0 & 5.0 & 2.5   \\
 V60H &20.0  &2.2  &100.0 &220.0  &45.5 &0.33  & 12.5 & 5.0 & 5.0  &5.0\\ 
 V97  &20.0  &6.0  &93.8 &562.8   &15.6 &1.52  & 10.0 & 5.0 & 5.0 & 2.5\\
\hline
\hline
\end{tabular}
\caption{Turbulent flame parameters for the DNS studies.}
\label{tbl:turb_param}
\end{table}

\section{Mathematical background} \label{sec:math}

\begin{figure}[h!]
\centering
\includegraphics[scale=0.50, trim=0.0cm 5.0cm 0.0cm 5.0cm]{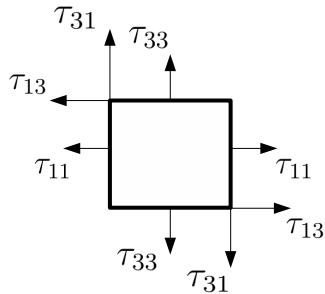}
\caption{Orientation of the stresses in the two non-homogeneous directions for this coordinate system.}
\label{fig:stresses_box}
\end{figure}

The filtered momentum equation in the case of a compressible flow reads, 

\begin{equation}
\ddxi{\bar{\rho} \tilde{u}_i}{t}+\ddxi{\bar{\rho}\tilde{u}_i\tilde{u}_j}{x_j}=-\ddxi{\bar{p}}{x_i}+\ddxi{{\tau}^r_{ij}}{x_j}-\ddxi{{\tau} _{ij} }{x_j}
\end{equation}

\noindent where the resolved and unresolved stress tensors ${\tau}_{ij}^r$ and ${\tau}_{ij}$ are given by,  

\begin{equation*}
{\tau}_{ij}^ {r}=\overline{\mu\left( \ddxi{u_i}{x_j}+\ddxi{u_j}{x_i}\right)}-\frac{2}{3}\delta _{ij}\overline{\mu \ddxi{u_k}{x_k}}
\end{equation*}

\noindent and

\begin{equation}
{\tau}_{ij}=\bar{\rho} \left( \widetilde{u_iu_j} -\tilde{u_i} \tilde{u_j}\right)
\end{equation}

\noindent respectively. The resolved stress tensor is typically approximated using the gradients of the filtered velocity components,

\begin{equation*}
\tau^r_{ij}\simeq \bar{\mu} \left( \ddxi{\tilde{u}_i}{x_j}+\ddxi{\tilde{u}_j}{x_i} -\frac{2}{3}\delta_{ij} \ddxi{\tilde{u}_k}{x_k} \right)=2\bar{\mu} \left( \tilde{S}_{ij} -\frac{1}{3}\delta_{ij}\tilde{S}_{kk} \right)
\end{equation*}

\noindent where 

\begin{equation*}
\tilde{S}_{ij}=\frac{1}{2} \left( \ddxi{\tilde{u}_i}{x_j}+\ddxi{\tilde{u}_j}{x_i} \right)
\end{equation*}

\noindent is the rate of strain tensor. $\tau_{ij}$ is an unclosed term and needs to be modelled in order to properly account for the effect of unresolved motions. Figure \ref{fig:stresses_box} shows the orientation of the normal and shear stresses for this particular configuration for the two non-homogeneous directions ($x-1$, $z-3$). In the case of turbulent premixed flames, a number of different models were evaluated for $\tau_{ij}$ a priori in \cite{klein_caf_2015}. The different models evaluated involved Smagorinsky, gradient, and three similarity-based models namely the similarity model of Bardina \cite{bardina_tf19_1983}, the similarity model of Vreman \cite{vreman_jfm_1997}, and an extended similarity model based on the work of Anderson and Domaratzki \cite{anderson_pof_2012}. Of all the models tested, the dynamic Smagorinsky model had the lowest correlation coefficient, while the remaining models were found to have about the same equally larger correlation coefficients. In comparison to the static version, the low correlation coefficients were found to be a result of the large spatial variations of the model's dynamic parameter $C_D$, which is a common problem with models of this kind. An alternative progress-variable based conditional averaging  regularisation procedure was proposed, however the performance of the dynamic Smagorinsky model was not found to improve substantially. Based on the findings in \cite{klein_caf_2015}, the models discussed in the text which follows were selected in order to benchmark against the ANN modelling approach. These include static and dynamic versions of the Smagorinsky model \cite{smagorinsky_mwr_1963}, two similarity-based models \cite{bardina_tf19_1983,klein_caf_2015}, the gradient and Clark models \cite{clark_jfm_1979}, and the WALE model \cite{nicoud_ftc_1999}. 

\subsection{Smagorinsky}

The Smagorinsky model is an eddy-diffusivity type of model \cite{moin_pof_1991}, 

\begin{equation}\label{eq:smag_model}
{\tau}_{ij} -\frac{1}{3}{\delta}_{ij} {\tau}_{kk} =-2 \bar{\rho} \nu _t \left( {\tilde{S}}_{ij}-\frac{1}{3} \delta _{ij} \tilde{S} _{kk} \right) 
\end{equation}

\noindent where the turbulent viscosity, $\nu _t$, is calculated using $\nu _t={C_D} \Delta ^2 |\tilde{S}|$, where $|\tilde{S}|=\sqrt{2\tilde{S}_{ij}\tilde{S}_{ij}}$. In the static version of the model, $C_D$ is replaced by $C_S ^2$ where $C_S \simeq 0.2$. The isotropic contribution (second term on left-hand side of Eq. \ref{eq:smag_model}) in many LES studies is typically absorbed into the filtered pressure term $\bar{p}$, assuming its contribution is relatively smaller. Erlebacher et al. \cite{erlebacher_jfm_1992} suggested that for relatively small SGS Mach numbers
the isotropic contribution of the stress tensor can indeed be neglected. Our focus in this study is on modelling the unresolved stress tensor regardless of the size of the isotropic part. As a result, the static Yoshizawa approximation is used in order to explicitly model the trace of the stress tensor \cite{yoshizawa_pof_86}, 

\begin{equation}
\tau _{kk}=2 \bar{\rho} C_I {\Delta}^2 |\tilde{S}|^2
\end{equation}

\noindent where $C_I=0.089$. Even though slightly different values were reported in the literature \cite{moin_pof_1991} this choice was found to produce quantitatively good results for the current DNS database, and is therefore employed throughout. The dynamic parameter $C_D$ is calculated following the least-squares approach suggested by Lilly \cite{lilly_pof_1992}, using \cite{salvetti_pof_1994}, 

\begin{equation}
C_D=\frac{ <-(L_{ij}-\frac{1}{3}\delta_{ij}L_{kk})M_{ij}>}{<2\Delta^2 M_{ij}M_{ij}>}
\end{equation}

\noindent where the Leonard term $L_{ij}$ is given by, 

\begin{equation*}
L_{ij}=\widehat{\bar{\rho}\tilde{u}_i\tilde{u}_j}-\widehat{(\bar{\rho}\tilde{u}_u)}\widehat{(\bar{\rho}\tilde{u}_j)} / \hat{\tilde{\rho}}
\end{equation*}

\noindent and,

\begin{equation}
M_{ij}=\alpha^2\hat{\bar{\rho}}|\hat{\tilde{S}}|\left( \hat{\tilde{S}}_{ij}-\frac{1}{3}\delta_{ij}\hat{\tilde{S}}_{kk} \right)-\left(  \widehat{ \bar{\rho}|\tilde{S}| \tilde{S}_{ij} }-\frac{1}{3}\delta_{ij}\widehat{ \bar{\rho}|\tilde{S}|S_{kk}} \right)
\end{equation}

The classic regualarisation approach is used in this study for $C_D$, as well as for other dynamic parameters in the models which folow, by volume-averaging ($< >$) in the homogeneous ($y$) direction \cite{moin_pof_1991}. Note that all test-filtering operations ( $\widehat{}$ ) are conducted at a filter width $\widehat{\Delta}=2\Delta$. 

\subsection{Scale-similarity} 

The scale-similarity model of Bardina (SIMB) \cite{bardina_tf19_1983} is given by,  

\begin{equation}
\tau _{ij}=\bar{\rho}(\overline{ \tilde{u}_i\tilde{u}_j}-\overline{\tilde{u}_i} \cdot \overline{\tilde{u}_j})
\end{equation}

This model showed improved predictions in comparison to the static Smagorinsky model for freely-propagating flames \cite{klein_caf_2015}. In \cite{klein_caf_2015} the following model based on the Inter-Scale Energy transfer model of Anderson and Domaratzki \cite{anderson_pof_2012} was also proposed (SIMET), 

\begin{equation}
\tau_{ij}=\bar{\rho} \left( \widehat{\hat{\tilde{u}}_i\tilde{u_j}} + \widehat{\hat{\tilde{u}}_j\tilde{u_i}}-\hat{\tilde{u}}_i\hat{\tilde{u}}_j- \widehat{\hat{\tilde{u}}_j\hat{\tilde{u}}_i} \right)
\end{equation}

\noindent which is Galilei invariant for the compressible case.  

\subsection{Gradient model}

The gradient model (GRAD) can be derived by expanding in Taylor series the filtered velocity product in the expression for $\tau_{ij}$ \cite{vreman_tcfd_1996} and retaining the leading term in the expansion \cite{clark_jfm_1979},

\begin{equation}
\tau _{ij}=\bar{\rho} \frac{\Delta ^2}{12}\ddxi{\tilde{u}_i}{x_k}\ddxi{\tilde{u}_j}{x_k}
\end{equation}

This model is known to perform well in a priori studies, and also forms the basis of many mixed models, such as the Clark model which is discussed next.

\subsection{Clark model}

Vreman et al. \cite{vreman_tcfd_1996} built upon the mixed model of Clark \cite{clark_jfm_1979} (CLARK) and produced the following dynamic mixed model with an eddy-diffusivity component complementing the gradient part in order to provide the necessary dissipation, 

\begin{equation}\label{eq:mdl_clark}
\tau _{ij}=\bar{\rho} \frac{\Delta ^2}{12}\ddxi{\tilde{u}_i}{x_k}\ddxi{\tilde{u}_j}{x_k}-C_C\bar{\rho} \Delta^2 |\tilde{S}'| \tilde{S'}_{ij}
\end{equation}
 
\noindent where, 

\begin{equation}
{S}'_{ij}(\underline{\tilde{u}})=\ddxi{\tilde{u}_i}{x_j}{}+\ddxi{\tilde{u}_j}{x_i}{}-\frac{2}{3}\delta_{ij}\ddxi{\tilde{u}_k}{x_k}{}=2\left( \tilde{S}_{ij}-\frac{1}{3}\delta_{ij}\tilde{S}_{kk} \right)
\end{equation}
 
\noindent and $|{S}'|=\left(0.5 {S}'_{ij}{S}'_{ij} \right)^{1/2}$. In the static version, the model parameter $C_C= 0.17^2$. In the dynamic version, it is calculated using,

\begin{equation}
C_C=\frac{<M_{ij}(L_{ij}-H_{ij})>}{<M_{ij}M_{ij}>}
\end{equation}

\noindent Denoting $v_i$=${\widehat{\bar{\rho}\tilde{u}_i}}/{\hat{\bar{\rho}}}$, the tensors $H_{ij}$ and $M_{ij}$ are given by, 

\begin{equation*}
H_{ij}=\hat{\bar{\rho}}\frac{\hat{\bar{\Delta}}^2}{12}\ddxi{v_i}{x_k}{}\ddxi{v_j}{x_k}{}
-\frac{\Delta^2}{12}{\left( \bar{\rho}\ddxi{\tilde{u}_i}{x_k}{}\ddxi{\tilde{u}_j}{x_k}{} \right)}^{\hat{}}
\end{equation*}

\noindent and

\begin{equation*}
M'_{ij}=-\hat{\bar{\rho}}{\hat{\bar{\Delta}}^2}|{S'}(\underline{v})|{S'}_{ij}(\underline{v})+\Delta^2{\left( \bar{\rho}|{S'}(\underline{\tilde{u}})|{S'}_{ij}(\underline{\tilde{u}}) \right)}^{\hat{}}
\end{equation*}

\noindent respectively. This model was evaluated both a priori and a posteriori, for the temporal mixing layer (non-reacting), with overall good results \cite{vreman_tcfd_1996}. In particular, in \cite{vreman_jfm_1997} this model was compared against the dynamic Smagorinsky, and the dynamic mixed model of Zhang et al. \cite{zhang_pof_1993}. The dynamic mixed model of Zhang et al. \cite{zhang_pof_1993} is of a similar nature but employs instead a scale-similarity term as the base model. The dynamic Clark model showed improved predictions over the dynamic Smagorinsky model, and equally good predictions to the scale-similarity based model \cite{zhang_pof_1993} albeit at a lower computational cost. Hence it serves as a good benchmark model in this study.

\subsection{WALE}

A similarity-based mixed model, based on the Wall-Adapting Local Eddy Viscosity (WALE) model \cite{nicoud_ftc_1999} was proposed in \cite{lodato_pof_2009}. This model was used to simulate a wall-impinging jet with overall good results \cite{lodato_pof_2009},

\begin{equation}\label{eq:wale}
\tau_{ij}-\frac{1}{3}\delta_{ij}\tau_{kk}=-2\bar{\rho}\nu_t \left( \tilde{S}_{ij}-\frac{1}{3}\delta_{ij}\tilde{S}_{kk} \right)+\bar{\rho}(\widehat{\tilde{u}_i\tilde{u}_j}-\hat{\tilde{u}}_i\hat{\tilde{u}}_j)
\end{equation}

\noindent This model has an eddy-diffusivity component like the Smagorinsky model, however the turbulent viscosity is calculated from the velocity gradient and shear rate tensors using, 

\begin{equation*}
\nu_t=(C_W\Delta)^2\frac{(\tilde{s}^d _{ij}\tilde{s}^d _{ij})^{3/2}}{(\tilde{S}_{ij}\tilde{S}_{ij})^{5/2}+(\tilde{s}^d _{ij}\tilde{s}^d _{ij})^{5/4}}
\end{equation*}

\noindent The model constant $C_W$=0.5, and $\tilde{s}^d _{ij}$ is the traceless symmetric part of the squared resolved velocity gradient tensor $\tilde{g}_{ij}=\partial{\tilde{u}_i}/\partial{x_j}$, 

\begin{equation*}
\tilde{s}^d _{ij}=\frac{1}{2} \left( \tilde{g}^2_{ij}+\tilde{g}^2_{ji} \right)-\frac{1}{3}\delta_{ij} \tilde{g}^2 _{kk}
\end{equation*}

\noindent where $g^2_{ij}=g_{ik}g_{kj}$. Note that in this case as well, the static Yoshizawa closure is used to model the trace of the stress tensor. 

\section{Filtering/sampling}

The DNS data are explicitly filtered on the fine DNS mesh using a Gaussian filter, $G(\underline{x})=\left( {{6}/{(\pi {\Delta} ^{2})}} \right)^{\frac{3}{2}}  \exp\left ( -{6\underline{x} \cdot \underline{x}}/{{\Delta} ^{2}}\right )$, where $\Delta$ is the corresponding filter width. The laminar flame thickness is used as a basis for filtering at $\Delta ^+=\Delta / \delta_L$=1.0, 2.0 and 3.0. Favre-filtered variables are defined as usual, $\tilde{\phi}(\underline{x},t)={\overline{\rho \phi}}/{\bar{\rho}}$. In order to simulate an LES, the filtered data are sampled onto a much coarser LES mesh with mesh spacing $h$, as indicated in Table \ref{tbl:mesh_les}. The LES mesh criterion developed in \cite{nikolaou_ftc_2018} is used, namely $h/\Delta$=1/4. This ensures that the filtered progress variable thickness is well-resolved on the coarse mesh. This results in LES meshes which are coarser than the Kolmogorov length scale of the incoming turbulence as indicated in Table \ref{tbl:mesh_les}: for cases V60 and V60H the ratio $h/\eta _k$ spans 3.6-10.8 while for case V97 it spans 7.8-23.3.  As a result, small-scale information of the order of $\eta_k$ is not resolved on the simulated LES mesh \cite{nikolaou_ftc_2018,nikolaou_prf_2018}. In contrast to most a priori studies in the literature which are conducted on the fine DNS mesh, this presents a more stringent a priori evaluation-gradients for example in classic models for the stress tensor discussed in the next section, are evaluated on the coarser LES mesh \cite{nikolaou_prf_2018}. The DNS data are filtered for a period of more than one flame time $t_{fl}=t/(\delta_L/s_L)$, and volume-averaged quantities have also been time-averaged in order to increase the statistical accuracy of the results.

\begin{table}
\centering
\begin{tabular}{l c c c c c}
\hline
\hline
Case & $\Delta ^+$ & $N_x$ & $N_y$ & $N_z$ & $h/\eta_k$\\ [0.75ex]
\hline
V60 &0   &513  &257  &257 &  \\  
    &1   &201  &101  &101 &3.6     \\
    &2   &101  &51   &51  &7.1    \\
    &3   &67   &34   &34  &10.8    \\
\hline
V60H &0  &641  &257  &257 &  \\
     &1  &251  &101  &101 &3.6     \\
     &2  &126  &51   &51  &7.1    \\
     &3  &83   &33   &33  &10.8    \\
\hline
V97 &0   &769  &385  &385 &  \\
    &1   &201  &101  &101 &7.8     \\
    &2   &101  &51   &51  &15.4    \\
    &3   &67   &34   &34  &23.3    \\
\hline
\hline
\end{tabular}
\caption{DNS and LES meshes for $h / \Delta$=1/4.}
\label{tbl:mesh_les}
\end{table}

\begin{figure}[h!]
\centering
\includegraphics[scale=0.35, trim=0.0cm 0.0cm 0.0cm 0.0cm]{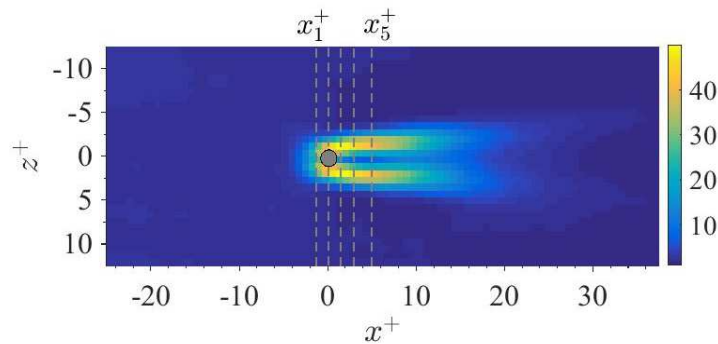}
\caption{Averaged (in homogeneous $y$ direction) normal stress component $\tau^+_{11}$, for case V60H and $\Delta^+=3$: averaging locations are shown in grey-dashed lines at $x^+_1$=-1.5 $\delta_L$, $x^+_2$=0, $x^+_3$=1.5 $\delta_L$, $x^+_4$=3 $\delta_L$, and $x^+_5$=5 $\delta_L$ relative to the rod location for all three DNS cases.}
\label{fig:avr_xi_highshear}
\end{figure}

Figure \ref{fig:avr_xi_highshear} shows the averaging locations in the streamwise direction along with the distribution of the averaged normal stress component $<\tau^+_{11}>$. Note that this component is symmetric around the rod as expected. The stresses are normalised ($^+$) using the unburnt mixture values $\rho_r$ and $s_L$. The contour plot in Fig. \ref{fig:avr_xi_highshear}  corresponds to case V60H, and the largest filter width $\Delta ^+$=3. The streamwise sampling locations are shown in dashed grey lines. These locations span a region of high shear immediately upstream and downstream of the rod at -1.5 $\delta_L$, 0 $\delta_L$, 1.5 $\delta_L$, 3 $\delta_L$ and 5 $\delta_L$ relative to the rod location. As one may observe from the results in Fig. \ref{fig:c_samp_avr_allcases} these locations also span regions with a high heat release, and are kept the same for all three DNS cases.

\section{Neural network training}\label{sec:ann_mdl}

The network is trained using the data on the coarse (LES) mesh hence the size of the training data reduces with increasing filter width. In order to train the network, a set of inputs and a set of outputs must first be identified. Gamahara and Hattori \cite{gamahara_prf_2017} used a three-layer network with each component of the stress tensor dealt with separately. Amongst the inputs tested, the individual velocity gradients were found to produce the best agreement against the DNS data. 
A range of possible inputs was also examined in \cite{wang_pof_2018} for modelling the unresolved stress tensor in incompressible homogeneous decaying turbulence. This involved combinations of the three velocity components $\bar{u}_i$ and their spatial first and second order derivatives \cite{wang_pof_2018}. In accordance with the results of Gamahara and Hattori \cite{gamahara_prf_2017} the velocity gradients were also found to be an important set of inputs. This is not surprising since many algebraic models for the stress tensor involve gradients of the velocity field.
In this study, the set of outputs involves the six unresolved stress-tensor components $\tau_{11}$, $\tau_{12}$, $\tau_{13}$, $\tau_{22}$, $\tau_{23}$, and $\tau_{33}$, for which predictions are required on the LES mesh. Based on the findings in previous studies \cite{gamahara_prf_2017,wang_pof_2018}, and after some experimentation with different sets of inputs, the best combination was found to be a set of ten variables as indicated in Table \ref{tbl:setio}. A similar choice of input variables was also found in \cite{gamahara_prf_2017,wang_pof_2018} not involving the density since the flow in those studies was incompressible. It is important to note that adding the velocity components $\tilde{u}_i$ to the sets of inputs in Table \ref{tbl:setio} deteriorated the predictions of the neural network. Note that the input and output variables are normalised as indicated in Table \ref{tbl:setio}.

\begin{table}[h!]
\centering
\begin{tabular}{l c }
\hline
\hline
 Inputs (10) & Outputs (6) \\ [0.75ex]
 \hline
$\bar{\rho}/\rho_r$ & $\tau_{11}/(\rho_r s^2_L)$ \\
$\partial{\tilde{u}}/ \partial{x_i}  \cdot \Delta /s_L$     & $\tau_{12}/(\rho_r s^2_L)$\\ 
$\partial{\tilde{v}}/ \partial{x_i} \cdot \Delta /s_L$     & $\tau_{13}/(\rho_r s^2_L)$\\ 
$\partial{\tilde{w}}/ \partial{x_i} \cdot \Delta /s_L$     & $\tau_{22}/(\rho_r s^2_L)$\\
                    & $\tau_{23}/(\rho_r s^2_L)$\\
                    & $\tau_{33}/(\rho_r s^2_L)$\\
\hline
\hline
\end{tabular}
\caption{Set of inputs and outputs for the neural network}
\label{tbl:setio}
\end{table}

\begin{figure}[h!]
\centering
\includegraphics[scale=0.45, trim=0.0cm 4.0cm 0.0cm 2.0cm]{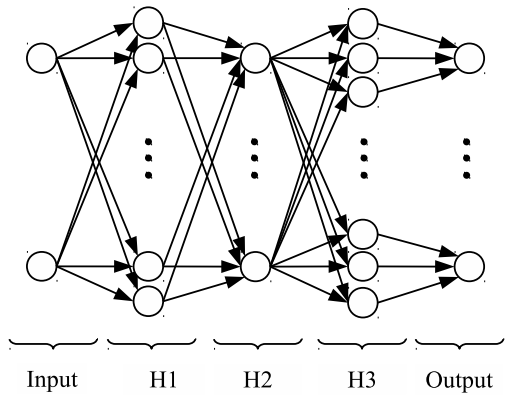}
\caption{Structure of the neural network-note that the last hidden layer is not fully connected to the output layer.}
\label{fig:nn_struct}
\end{figure}

The network structure is shown in Fig. \ref{fig:nn_struct}. The input layer consists of the set of 10 input variables. The first hidden layer (H1) consists of 40 nodes, the second hidden layer (H2) has 10 nodes, and the last hidden layer (H3) consists of 18 nodes. The output layer consists of 6 nodes, one for each stress tensor component. Rectified Linear Unit (RELU) activation functions are used in the three hidden layers, and a linear activation function in the output layer. All layers but the output layer are fully connected. In the output layer, a set of three nodes only are connected to each output node, as shown in Fig. \ref{fig:nn_struct}. This choice was found to improve the predictions for individual stress components, whose magnitude for this flow configuration differs substantially. As a result, the weights of the three nodes in the third hidden layer  are allowed in this structure to assume independent values in order to reflect the relatively large difference in magnitude amongst the six stress tensor components. 

Table \ref{tbl:train_test} shows the training/testing scenario used in the present study. Cases V60 and V60H have lower turbulence levels in comparison to case V97, and the flame structure is less convolved as opposed to V97 as one may observe from Figs. \ref{fig:c_01_iso} and \ref{fig:c_samp_avr_allcases}. Therefore the choice of this training/testing scenario is a demaning one since the network is required to make predictions for the shear stresses on a flame having a much more convolved and patchy flame surface. The training and testing is conducted for each individual filter width, with the network structure being kept the same between the different filter widths. This results in three different networks (one for each $\Delta^+$) with the same structure, but having different node weights in each case. Around 90\% of the data from cases V60 and V60H are used for training and 10\% for validation. A mean-squared error is used as an objective function, and 1000 epochs as the maximum number of iterations. The weights are chosen at the point where the validation error is minimised. The open-source Python-based library ``TensorFlow" was used for training the network \cite{tensor_flow}. 

\begin{table}
\centering
\begin{tabular}{l c c}
\hline
\hline
 Case & Training/validation & Testing \\ [0.75ex]
 \hline
V60   &   x     & - \\
V60H  &   x     & - \\
V97   &   -     & x \\
\hline
\hline
\end{tabular}
\caption{Training and testing scenario.}
\label{tbl:train_test}
\end{table}

\section{Model performance evaluation}\label{sec:mdl_eval}

\begin{figure}[h!]
\centering
\includegraphics[scale=0.65, trim=0.0cm 0.0cm 0.0cm 0.0cm]{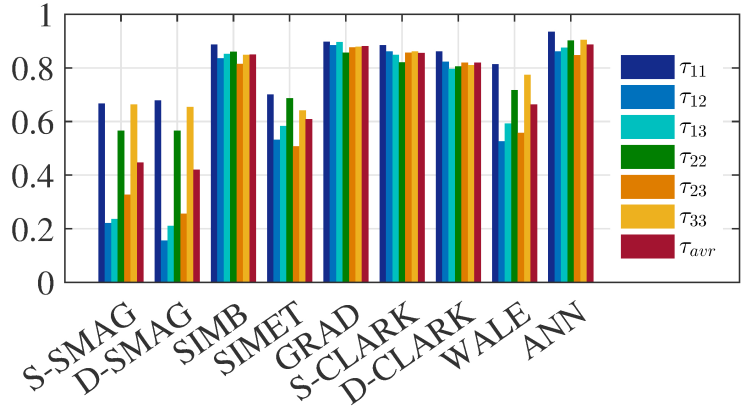}
\caption{Pearson correlation coefficients averaged across all filter widths, for each stress tensor component, for case V97. }
\label{fig:R_str_avr_delta_V97}
\end{figure}

In order to evaluate the performance of the models, Pearson correlation coefficients as well as spatial averages have been calculated. The Pearson coefficients between the modelled and actual stresses on the LES mesh, averaged across all three filter widths, are shown in Fig. \ref{fig:R_str_avr_delta_V97} for each stress tensor component. The Smagorinsky model has the lowest correlation coefficients both for its static and dynamic version, in accordance with the results in \cite{klein_caf_2015}. The similarty, gradient and Clark models have some of the highest correlation coefficients, which is typical in a priori tests of these models. The ANN-based model also has some of the highest correlation coefficients for all six components of the stress tensor.

\begin{figure}[h!]
\centering
\subfigure[]{
\includegraphics[scale=0.7, trim=2.0cm 1.0cm 0.0cm 0.0cm]{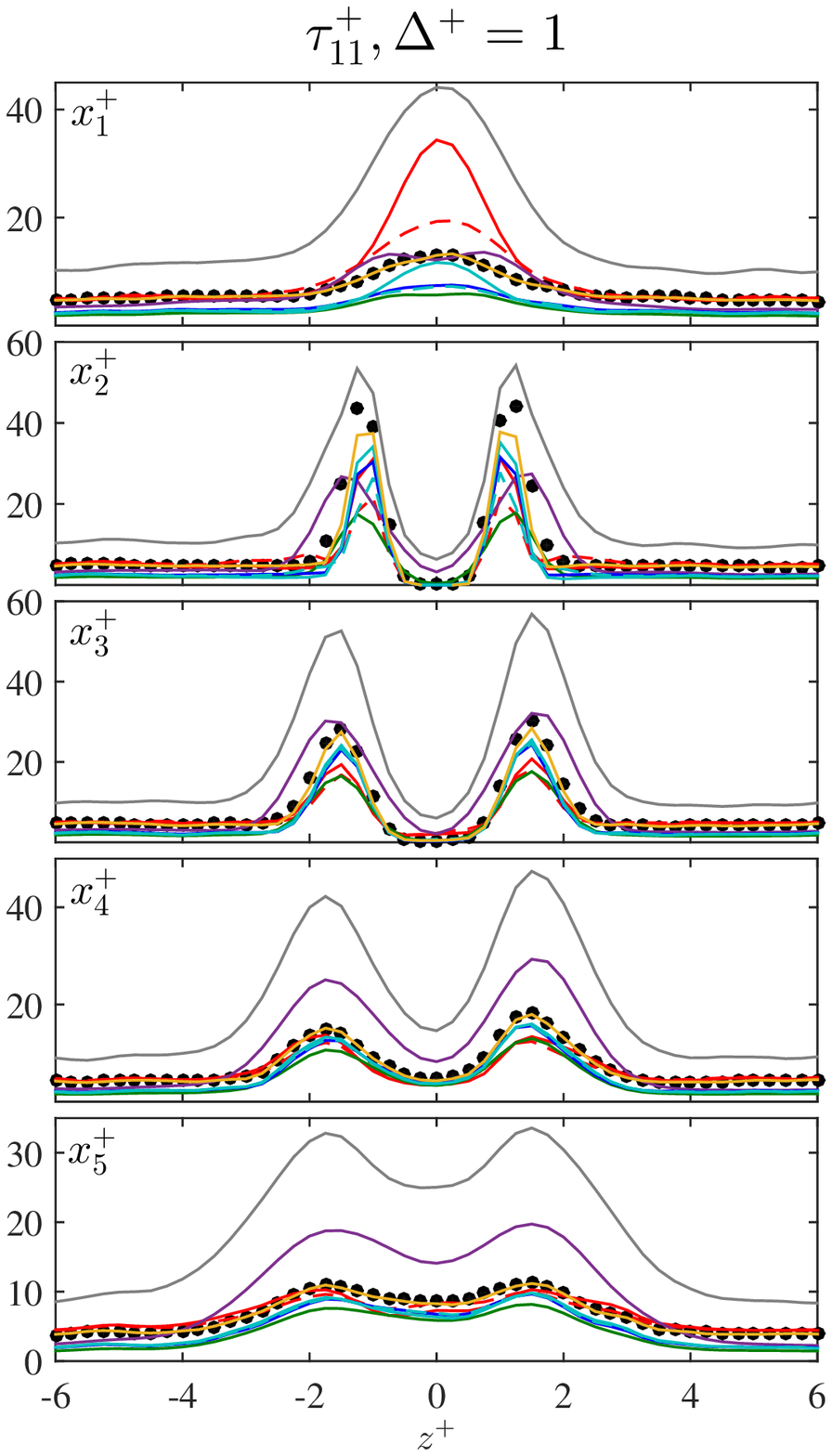}
}
\subfigure[]{
\includegraphics[scale=0.7, trim=1.0cm 1.0cm 0.0cm 0.0cm]{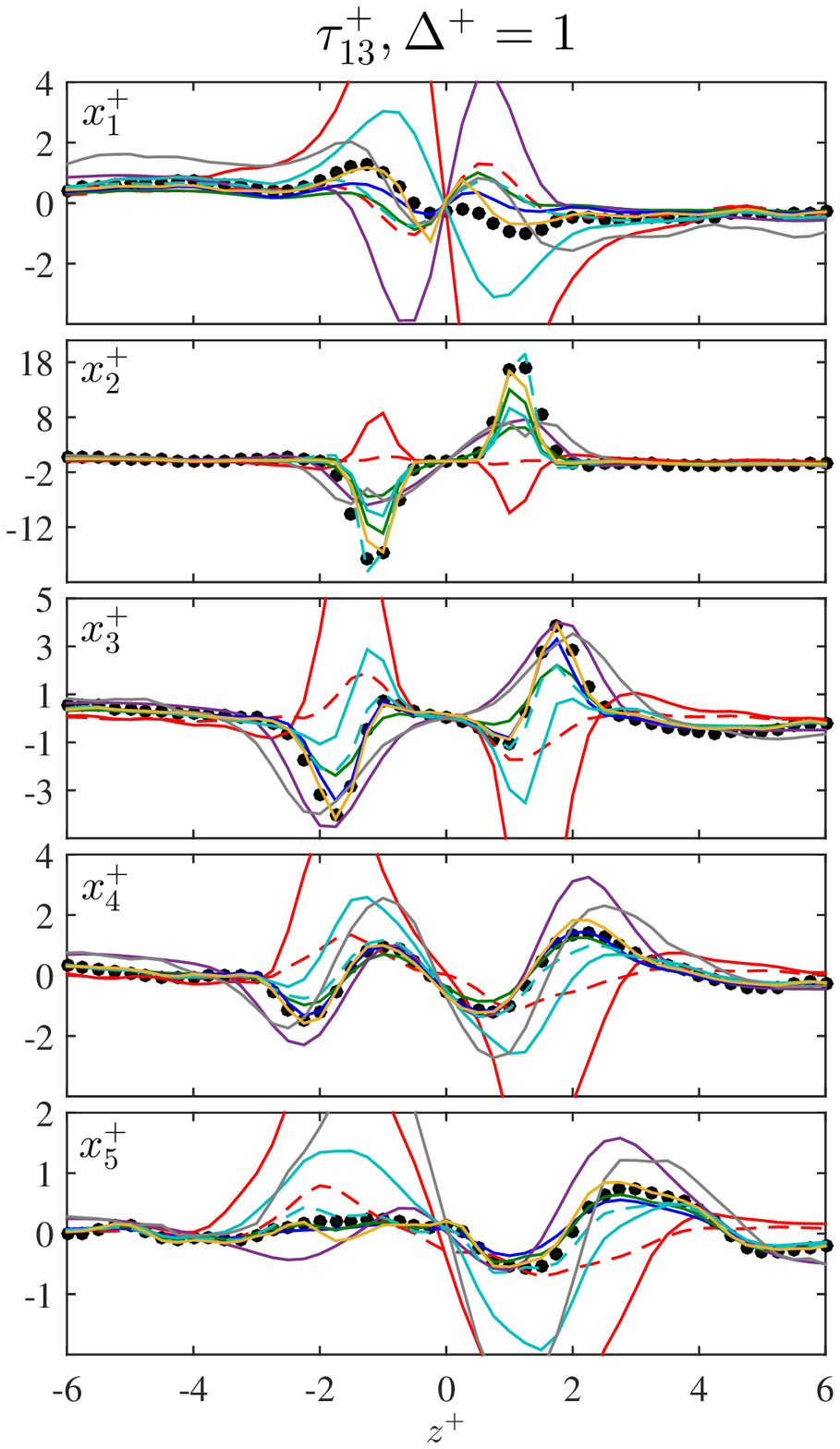}
}
\caption{Average shear stresses (a) $<\tau^+_{11}>$ and (b) $<\tau^+_{13}>$, for different $x$ locations, case V97, $\Delta^+=1.0$.  \protect\redline $S-SMAG$, \protect\redlinedash $D-SMAG$ \protect\blueline $SIMB$, \protect\purpleline $SIMET$, \protect\greenline $GRAD$, \protect\cyanline $S-CLARK$, \protect\cyanlinedash $D-CLARK$, \protect\greyline $WALE$, \protect\yellowline $ANN$. }
\label{fig:str_avr_1113_V97_010}
\end{figure}

\begin{figure}[h!]
\centering
\subfigure[]{
\includegraphics[scale=0.7, trim=2.0cm 1.0cm 0.0cm 0.0cm]{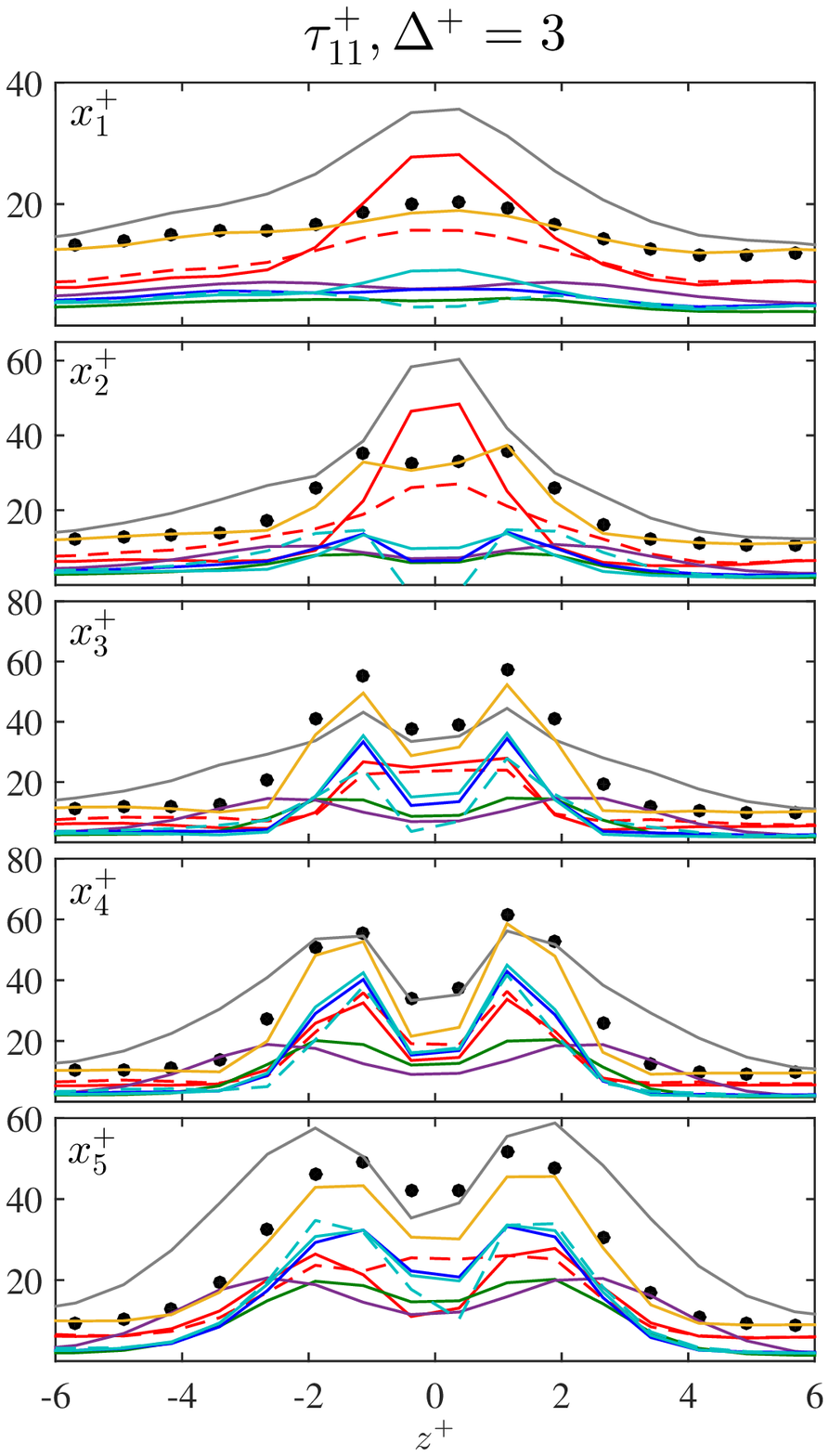}
}
\subfigure[]{
\includegraphics[scale=0.7, trim=1.0cm 1.0cm 0.0cm 0.0cm]{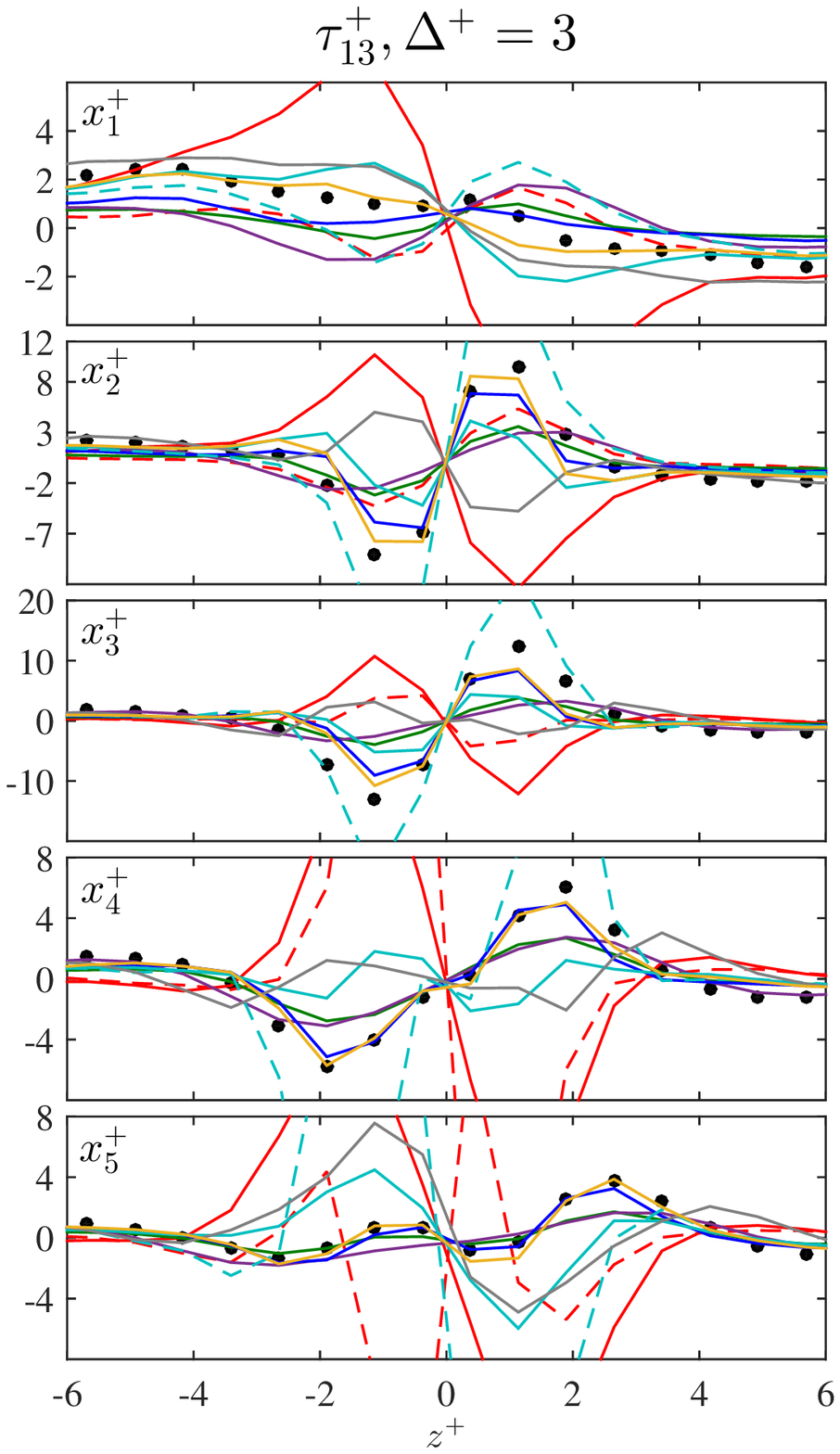}
}
\caption{Average shear stresses (a) $<\tau^+_{11}>$ and (b) $<\tau^+_{13}>$, for different $x$ locations, case V97, $\Delta^+=3.0$. Lines as in Fig. \ref{fig:str_avr_1113_V97_010}.}
\label{fig:str_avr_1113_V97_030}
\end{figure}

Figures \ref{fig:str_avr_1113_V97_010} and \ref{fig:str_avr_1113_V97_030} show for the smallest and largest filter widths respectively,
the average (in $y$) variation of the largest stress tensor components for this flow configuration namely $\tau^+_{11}$ and $\tau^+_{13}$. Note that these are normalised using the laminar flame values. For the classic models significant variations are observed even for the smallest filter width across all streamise sampling locations. The Smagorinsky model fails to predict the correct variation of $\tau_{13}$ for $\Delta^+=1$ and the predictions for $\Delta^+=3$ deteriorate even more. The SIMB and GRAD models provide reasonable predictions for $\Delta^+=1$, however their predictions also deteriorate somewhat for $\Delta^+=3$. The ANN-based model performs reasonably well outperforming the rest of the models, for both filter widths, and for all five streamwise sampling locations. This is also reflected in the high correlation coefficients obtained for this model in Fig. \ref{fig:R_str_avr_delta_V97}.

It is important to note at this point that the network developed in this study will perform reasonably well for conditions not too dissimilar to those found in the current database. This is typical of all data-driven methods. In principle, networks of enhanced generality/accuracy can be obtained by training on larger and more diverse data sets, spanning different flow configurations and reaction modes. 
 
%

\clearpage

\section*{Conclusions}

The application of artificial neural networks for modelling the unresolved stress tensor in a highly demanding multi-physics flow configuration including turbulence, reaction, and mean shear
is investigated a priori, using data from a high-fidelity direct numerical simulation database of a rod-stabilised hydrogen-air turbulent premixed V-flame.

In contrast to most a priori studies in the literature, the evaluation of this modelling approach is conducted following a simulated LES approach, where the data are first filtered on the fine mesh, and then sampled onto the coarser LES mesh. This ensures that mesh-coarsening effects are accounted for. Suitable inputs are identified for the neural network, which include the filtered density, and the gradients of the filtered velocity components. The network output consists of all six individual components of the stress tensor. The performance of the neural network-based modelling approach is compared against the performance of eight other classic algebraic models in the literature namely
Smagorinsky, scale-similarity, gradient/Clark, and models of the mixed kind, including both static and dynamic formulations. The application of the neural network is demonstrated with success, despite the highly demanding flow configuration, with large correlation coefficients and good agreement of spatial averages against the DNS data, for all six components of the stress tensor.

Data-driven modelling approaches for such highly non-linear multi-physics flows, provide a straightforward way of obtaining best-fit functions where closed-form algebraic solutions are otherwise difficult to obtain. In principle, networks of enhanced generality can be constructed by employing a wider range of training datasets and/or number and type of input variables. 

\vspace{0.5cm}

 
 \clearpage
 


\begin{thebibliography}{10}

\bibitem{pitsch_annrev_2006}
H.~Pitsch, ``Large {E}ddy simulation of turbulent combustion,'' {\em Ann. Rev.
  Fluid Mech.}, vol.~38, pp.~453--482, 2006.

\bibitem{gicquel_pec_2012}
L.~Y.~M. Gicquel, G.~Staffelbach, and T.~Poinsot, ``Large {E}ddy simulations of
  gaseous flames in gas turbine combustion chambers,'' {\em Prog. En. Combust.
  Sc.}, vol.~38, pp.~782--817, 2012.

\bibitem{sagaut_book_2001}
P.~Sagaut, ``Large {E}ddy {S}imulation for {I}ncompressible {F}lows: {A}n
  {I}ntroduction,'' {\em Springer-Verlag}, 2001.

\bibitem{meneveau_annrev_2000}
C.~Meneveau and J.~Katz, ``Scale invariance and turbulence models for
  large-eddy simulation,'' {\em Ann. Rev. Fluid. Mech.}, vol.~32, pp.~1--32,
  2000.

\bibitem{smagorinsky_mwr_1963}
J.~Smagorinsky, ``General circulation experiments with the primitive
  equations,'' {\em {M}onthly {W}eath. {R}ev.}, vol.~91, pp.~99--164, 1963.

\bibitem{germano_pof_1991}
M.~Germano, U.~Piomelli, P.~Moin, and W.~Cabot, ``A dynamic sub-grid scale eddy
  viscosity model,'' {\em Phys. Fluids}, vol.~3, pp.~1760--1765, 1991.

\bibitem{moin_pof_1991}
P.~Moin, K.~Squires, W.~Cabot, and S.~Lee, ``A dynamic sub-grid scale model for
  compressible turbulence and scalar transport,'' {\em J. Fluid Mech.}, vol.~3,
  pp.~2746--2757, 1991.

\bibitem{bardina_tf19_1983}
J.~Bardina, J.~Ferziger, and W.~Reynolds, ``Improved turbulence models based on
  large eddy simulation of homogeneous, incompressible, turbulent flows,'' {\em
  Technical Report no. TF-19, Dep. Mech. Eng. Stanford University}, 1983.

\bibitem{vreman_tcfd_1996}
B.~Vreman, B.~Geurts, and H.~Kuerten, ``Large eddy simulation of the temporal
  mixing layer using the {C}lark model,'' {\em Theoret. Comput. Fluid
  Dynamics}, vol.~8, pp.~309--324, 1996.

\bibitem{clark_jfm_1979}
R.~A. Clark, ``Evaluation of sub-grid scalar models using an accurately
  simulated turbulent flow,'' {\em J. Fluid Mech.}, vol.~91, pp.~1--16, 1979.

\bibitem{zhang_pof_1993}
Y.~Zhang, R.~Street, and J.~Koseff, ``A dynamic mixed subgrid-scale model and
  its application to turbulent recirculating flows,'' {\em Phys. Fluids},
  vol.~5, pp.~3186--3196, 1993.

\bibitem{salvetti_pof_1994}
M.~Salvetti, ``A priori tests of a new dynamic sub-grid scale model for finite
  difference large eddy simulations,'' {\em Phys. Fluids}, vol.~7,
  pp.~2831--2847, 1994.

\bibitem{nicoud_ftc_1999}
F.~Nicoud and F.~Ducros, ``Sub-grid scale stress modelling based on the square
  of the velocity gradient tensor,'' {\em Flow Turb. Combust.}, vol.~62,
  pp.~183--200, 1999.

\bibitem{lodato_pof_2009}
G.~Lodato, L.~Vervisch, and P.~Domingo, ``A compressible wall-adapting
  similarity mixed model for large-eddy simulation of the impinging round
  jet,'' {\em Phys. Fluids}, vol.~21, pp.~1--21, 2009.

\bibitem{wincelmans_pof_2001}
G.~Wincelmans, A.~Wray, O.~Vasilyev, and H.~Jeanmart, ``Explicit-filtering
  large-eddy simulation using the tensor-diffusivity model supplemented by a
  dynamic smagorinsky term,'' {\em Phys. Fluids}, vol.~13, pp.~1385--1403,
  2001.

\bibitem{tao_pof_2000}
B.~Tao, J.~Katz, and C.~Meneveau, ``Geometry and scale relationships in high
  {R}eynolds number turbulence determined from three-dimensional holographic
  velocimetry,'' {\em Phys. Fluids}, vol.~12, pp.~941--944, 2000.

\bibitem{tao_jfm_2002}
B.~Tao, J.~Katz, and C.~Meneveau, ``Statistical geometry of subgrid-scale
  stresses determined from holographic velocimetry measurements,'' {\em J.
  Fluid Mech.}, vol.~457, pp.~35--78, 2002.

\bibitem{pfandler_expfluids_2010}
P.~Pfandler, F.~Beyrau, F.~Dinkelacker, and A.~Leipertz, ``A prior testing of
  an eddy viscosity model for the density-weighted sub-grid scale stress tensor
  in turbulent premixed flames,'' {\em Exp. fluids}, vol.~49, pp.~839--851,
  2010.

\bibitem{klein_caf_2015}
M.~Klein, C.~Kasten, Y.~Gao, and N.~Chakraborty, ``A priori direct numerical
  simulation assessment of sub-grid scale stress tensor closures for turbulent
  premixed combustion,'' {\em Comp. Fluids}, vol.~122, pp.~1--1, 2015.

\bibitem{krizhevsky_procneural_2012}
A.~Krizhevsky, I.~Sutskever, and G.~Hinton, ``Imagenet classification with deep
  convolutional neural networks,'' {\em Proc. Advances in Neural Information
  Processing Systems}, vol.~25, pp.~1090--1098, 2012.

\bibitem{sutskever_procadvneur_2014}
I.~Sutskever, O.~Vinyals, and Q.~Le, ``Sequence to sequence learning with
  neural networks,'' {\em Proc. Advances in Neural Information Processing
  Systems}, vol.~27, pp.~3104--3112, 2014.

\bibitem{Mnih_nat_2015}
V.~Mnih, K.~Kavukcuoglu, D.~Silver, A.~Rusu, J.~Veness, M.~Bellemare,
  A.~Graves, M.~Riedmiller, A.~Fidjeland, G.~Ostrovski, S.~Petersen,
  C.~Beattie, A.~Sadik, I.~Antonoglou, H.~King, D.~Kumaran, D.~Wierstra,
  S.~Legg, and D.~Hassabis, ``Human-level control through deep reinforcement
  learning,'' {\em Nature}, vol.~518, pp.~529--533, 2015.

\bibitem{silver_nat_2016}
D.~Silver, ``Mastering the game of go with deep neural networks and tree
  search,'' {\em Nature}, vol.~529, pp.~484--489, 2016.

\bibitem{Khan_nat_2001}
J.~Khan, J.~Wei, M.~Ringer, L.~Saal, M.~Ladanyi, F.~Westermann, F.~Berthold,
  M.~Schwab, C.~Antonescu, C.~Peterson, and P.~Meltzer, ``Classification and
  diagnostic prediction of cancers using gene expression profiling and
  artificial neural networks,'' {\em Nature}, vol.~7, pp.~673--679, 2001.

\bibitem{kanov_compsceng_2015}
K.~Kanov, R.~Burns, C.~Lalescu, and G.~Eyink, ``The johns hopkins turbulence
  databases: an open simulation laboratory for turbulence research,'' {\em
  Comput. Sci. Eng.}, vol.~17, pp.~10--17, 2015.

\bibitem{minamoto_pof_2011}
Y.~Minamoto, N.~Fukushima, M.~Tanahashi, T.~Miyauchi, and T.~Dunstan, ``Effect
  of flow geometry on turbulence-scalar interaction in premixed flames,'' {\em
  Phys. Fluids}, vol.~23, pp.~1--18, 2011.

\bibitem{nikolaou_cnf_2014}
Z.~M. Nikolaou and N.~Swaminathan, ``Evaluation of a reduced mechanism for
  turbulent premixed combustion,'' {\em Combust. Flame}, vol.~161,
  pp.~3085--3099, 2014.

\bibitem{nikolaou_cst_2015}
Z.~M. Nikolaou and N.~Swaminathan, ``Direct numerical simulation of complex
  fuel combustion with detailed chemistry: physical insight and mean reaction
  rate modelling,'' {\em Comb. Sc. Tech.,}, vol.~187, pp.~1759--1789, 2015.

\bibitem{aspden_cnf_2016}
A.~Aspden, M.~Day, and J.~Bell, ``Three-dimensional direct numerical simulation
  of turbulent lean premixed methane combustion with detailed kinetics,'' {\em
  Combust. Flame}, vol.~166, pp.~266--283, 2016.

\bibitem{wang_jfm_2017}
H.~Wang, E.~Hawkes, J.~Chen, and B.~Zhou, ``Direct numerical simulations of a
  high {K}arlovitz number laboratory premixed jet flame – an analysis of
  flame stretch and flame thickening,'' {\em J. Fluid Mech.}, vol.~815,
  pp.~511--536, 2017.

\bibitem{jchen_exapp_2017}
S.~Treichler, M.~Bauer, A.~Bhagatwala, G.~Borghesi, R.~Sankaran, P.~M.
  H.~Kolla, E.~Slaughter, W.~Lee, A.~Aiken, and J.~Chen, ``{S3D}-{L}egion: {A}n
  {E}xascale {S}oftware for {D}irect {N}umerical {S}imulation of {T}urbulent
  {C}ombustion with {C}omplex {M}ulticomponent {C}hemistry,'' {\em Exasc.
  Scient. Appl.}, vol.~12, pp.~257--258, 2017.

\bibitem{hernandez_caf_2018}
F.~H. Perez, N.~Mukhadiyev, X.~Xu, A.~Sow, B.~Li, R.~Sankaran, and H.~Im,
  ``Direct numerical simulation of reacting flows with detailed chemistry using
  many-core cpu acceleration,'' {\em Comp. Fluids}, vol.~173, pp.~73--79, 2018.

\bibitem{kutz_jfm_2017}
J.~Kutz, ``Deep learning in fluid dynamics,'' {\em J. Fluid Mech.}, vol.~814,
  pp.~1--4, 2017.

\bibitem{hornik_nn_1991}
K.~Hornik, ``Approximation capabilities of multi-layer feedforward networks,''
  {\em Neural networks}, vol.~4, pp.~251--257, 1991.

\bibitem{milano_jcp_2002}
M.~Milano and P.~Koumoutsakos, ``Neural network modelling for near wall
  turbulent flow,'' {\em J. Comput. Phys.}, vol.~182, pp.~1--26, 2002.

\bibitem{ling_jfm_2016}
J.~Ling, A.~Kurawski, and J.~Templeton, ``Reynolds averaged turbulence
  modelling using deep neural networks with embedded invariance,'' {\em J.
  Fluid Mech.}, vol.~807, pp.~155--166, 2016.

\bibitem{tracy_aiaa_2015}
B.~Tracy, K.~Duraisamy, and J.~Alonson, ``A machine learning stratety to assist
  turbulence model development,'' {\em 53rd {AIAA} {A}erospace {S}ciences
  {M}eeting, {AIAA} {S}ci{T}ech {F}orum}, pp.~1--22, 2015.

\bibitem{wang_prf_2017}
J.~Wang, J.~Wu, and H.~Xiao, ``Physics-informed machine learning approach for
  reconstructing reynolds stress modelling discrepancies based on {DNS} data,''
  {\em Phys. Rev. Fluids}, vol.~2, p.~034603, 2017.

\bibitem{singh_AIAA_2017}
A.~Singh, S.~Melida, and K.~Duraisamy, ``Machine-learning-augmented predictive
  modeling of turbulent separated flows over airfoils,'' {\em AIAA}, vol.~55,
  pp.~2215--2227, 2017.

\bibitem{ma_pof_2015}
M.~Ma, J.~Lu, and G.~Tryggvason, ``Using statistical learning to close two-
  fluid multiphase flow equations for a simple bubbly system,'' {\em Phys.
  Fluids}, vol.~27, p.~092101, 2015.

\bibitem{sarghini_caf_2003}
F.~Sarghini, G.~de~Felice, and S.~Santini, ``Neural networks based subgrid
  scale modeling in large eddy simulations,'' {\em Comp. Fluids}, vol.~97-108,
  p.~32, 2003.

\bibitem{moreau_pof_2006}
A.~Moreau, O.~Teytaud, and J.~Bertoglio, ``Optimal estimation for large-eddy
  simulation of turbulence and application to the analysis of subgrid models,''
  {\em Phys. Fluids}, vol.~18, p.~105101, 2006.

\bibitem{gamahara_prf_2017}
M.~Gamahara and Y.~Hattori, ``Searching for turbulence models by artificial
  neural network,'' {\em Phys. Rev. Fluids}, vol.~2, p.~054604, 2017.

\bibitem{maulik_jfm_2017}
R.~Maulik and O.~San, ``A neural network approach for the blind deconvolution
  of turbulent flows,'' {\em J. Fluid Mech.}, vol.~831, pp.~151--181, 2017.

\bibitem{wang_pof_2018}
Z.~Wang, K.~Luo, D.~Li, J.~Tan, and J.~Fan, ``Investigations of data-driven
  closure for subgrid-scale stress in large-eddy simulation,'' {\em Phys.
  Fluids}, vol.~30, p.~125101, 2018.

\bibitem{ihme_proccomb_2009}
M.~Ihme, C.~Schmitt, and H.~Pitsch, ``Optimal artificial neural networks and
  tabulation methods for chemistry representation in {LES} of a bluff-body
  swirl-stabilized flame,'' {\em Proc. Combust. Inst.}, vol.~32,
  pp.~1527--1535, 2009.

\bibitem{sen_proccomb_2009}
B.~Sen and S.~Menon, ``Turbulent premixed flame modeling using artificial
  neural networks based chemical kinetics,'' {\em Proc. Comb. Inst.}, vol.~32,
  pp.~1605--1611, 2009.

\bibitem{sen_cnf_2010}
B.~Sen, E.~Hawkes, and S.~Menon, ``Large eddy simulation of extinction and
  reignition with artificial neural networks based chemical kinetics,'' {\em
  Combust. Flame}, vol.~157, pp.~566--578, 2010.

\bibitem{lapeyre_arxive_2018}
C.~Lapeyre, A.~Misdariis, N.~Cazard, D.~Veynante, and T.~Poinsot, ``Training
  convolutional neural networks to estimate turbulent sub-grid scale reaction
  rates,'' {\em arXiv:1810.03691 [physics.flu-dyn]}, 2018.

\bibitem{nikolaou_arxive_2018}
Z.~M. Nikolaou, C.~Chrysostomou, L.~Vervisch, and S.~Cant, ``Modelling
  turbulent premixed flames using convolutional neural networks: application to
  sub-grid scale variance and filtered reaction rate,'' {\em arXiv:1810.07944
  [physics.flu-dyn]}, 2018.

\bibitem{nikolaou_ftc_2019}
Z.~M. Nikolaou, C.~Chrysostomou, L.~Vervisch, and S.~Cant, ``Progress variable
  variance and filtered rate modelling using convolutional neural networks and
  flamelet methods,'' {\em Flow Turb. Combust. In press}, 2019.

\bibitem{bray_cst_1981}
K.~Bray, P.~Libby, G.~Masuya, and J.~Moss, ``Turbulence production in premixed
  turbulent flames,'' {\em Combust. Sc. Technol.}, vol.~25, pp.~127--140, 1981.

\bibitem{chomiak_cnf_1995}
J.~Chomiak and J.~Nisbet, ``Modelling variable density effects in turbulent
  flames-some basic considerations,'' {\em Combust. Flame}, vol.~102,
  pp.~371--386, 1995.

\bibitem{schoepplein_jcp_2018}
M.~Schoepplein, J.~Weatheritt, R.~Sandberg, M.~Talei, and M.~Klein,
  ``Application of an evolutionary algorithm to {LES} modelling of turbulent
  transport in premixed flames,'' {\em Journal of Computational Physics},
  vol.~374, pp.~1166 -- 1179, 2018.

\bibitem{nikolaou_ftc_2018}
Z.~M. Nikolaou and L.~Vervisch, ``A priori assessment of an iterative
  deconvolution method for {LES} sub-grid scale variance modelling,'' {\em Flow
  Turb. Combust.}, vol.~101, pp.~33--53, 2018.

\bibitem{nikolaou_prf_2018}
Z.~M. Nikolaou, L.~Vervisch, and R.~S. Cant, ``Scalar flux modelling in
  turbulent flames using iterative deconvolution,'' {\em Phys. Rev. Fluids},
  vol.~3, p.~043201, 2018.

\bibitem{Gutheil1993}
E.~Gutheil, G.~Balakrishnan, and F.~A. Williams, ``Structure and {E}xtinction
  of {H}ydrogen-{A}ir {D}iffusion {F}lames,'' in {\em Lecture Notes in Physics:
  Reduced kinetic mechanisms for applications in combustion systems.}
  (N.~Peters and B.~Rogg, eds.), pp.~177--195, New York: Springer Verlag, 1993.

\bibitem{kee_sandia_86}
R.~J. Kee, G.~Dixon-Lewis, J.~Warnatz, M.~E. Coltrin, and J.~A. Miller, ``A
  {F}ortran computer code package for the evaluation of gas-phase
  multicomponent transport properties,'' {\em Report No. SAND86-8246, Sandia
  National Laboratories, Livermore, CA, USA}, 1986.

\bibitem{kee_sandia_89}
R.~J. Kee, F.~M. Rupley, and J.~A. Miller, ``Chemkin-{II}: {A F}ortran chemical
  kinetics package for the analysis of gas phase chemical kinetics,'' {\em
  Report No. SAND89-8009B, Sandia National Laboratories, Livermore, CA, USA,
  1989}, 1989.

\bibitem{vreman_jfm_1997}
B.~Vreman, B.~Geurts, and H.~Kuerten, ``Large-eddy simulation of the temporal
  mixing layer,'' {\em J. Fluid Mech.}, vol.~339, pp.~357--390, 1997.

\bibitem{anderson_pof_2012}
B.~Anderson and J.~Domaradzki, ``A subgrid-scale model for large-eddy
  simulation based on the physics of interscale energy transfer in
  turbulence,'' {\em Phys. Fluids}, vol.~24, pp.~1--35, 2012.

\bibitem{erlebacher_jfm_1992}
G.~Erlebacher, M.~Hussaini, C.~Speziale, and T.~Zang, ``Toward the large-eddy
  simula- tion of compressible turbulent flows,'' {\em J. Fluid Mech.},
  vol.~238, pp.~155--185, 1992.

\bibitem{yoshizawa_pof_86}
A.~Yoshizawa, ``Statistical theory for compressible turbulent shear flows, with
  the application to sub-grid modelling,'' {\em Phys. Fluids}, vol.~29,
  pp.~2152--2164, 1986.

\bibitem{lilly_pof_1992}
D.~Lilly, ``A proposed modification of the {G}ermano subgrid-scale closure
  method,'' {\em Phys. Fluids}, vol.~4, pp.~633--635, 1992.

\bibitem{tensor_flow}
M.~Abadi, A.~Agarwal, P.~Barham, E.~Brevdo, Z.~Chen, C.~Citro, G.~S. Corrado,
  A.~Davis, J.~Dean, M.~Devin, S.~Ghemawat, I.~Goodfellow, A.~Harp, G.~Irving,
  M.~Isard, Y.~Jia, R.~Jozefowicz, L.~Kaiser, M.~Kudlur, J.~Levenberg,
  D.~Man\'{e}, R.~Monga, S.~Moore, D.~Murray, C.~Olah, M.~Schuster, J.~Shlens,
  B.~Steiner, I.~Sutskever, K.~Talwar, P.~Tucker, V.~Vanhoucke, V.~Vasudevan,
  F.~Vi\'{e}gas, O.~Vinyals, P.~Warden, M.~Wattenberg, M.~Wicke, Y.~Yu, and
  X.~Zheng, ``{TensorFlow}: Large-scale machine learning on heterogeneous
  systems,'' 2015.

\end{thebibliography}


\clearpage
\newpage

\end{document}